# Performance evaluation of a novel relay assisted hybrid FSO / RF communication system with receive diversity

*Mohammad Ali Amirabadi[1],✉*
[1] *School of Electrical Engineering, Iran University of Science and Technology, Tehran, Iran*
✉ *E-mail: m_amirabadi@elec.iust.ac.ir*

**Abstract:** One of the main problems in mobile communication systems is the degradation of Radio Frequency (RF) connection when mobile user is far from base station. One way to solve this problem is to increase the transmitter power, but the mobile transmitter is not able to supply much power. Another way is to use a relay; among relay schemes, amplify and forward is better for long range communications. Amplify and forward relay is not affordable in terms of power consumption and performance, because it consumes a lot of power inefficiently and enhances the noise. Therefore, in other cases, except in the case of long range links, other relay protocols, such as decode and forward, as well as demodulate and forward, are preferable. In this paper, a novel multi-hop hybrid Free Space Optical (FSO) / RF link is presented; it is made up of two main parts. The first part establishes the connection between the mobile user and source base station, and the second part establishes the connection between the source and the destination base stations. In the first part, a mobile user wants to connect to the source base station via a long range link; therefore, a fixed gain amplify and forward relay with multiple receive antennas is used for communication establishment. In the second part, the source and the destination base stations are connected via a multi-hop hybrid parallel FSO / RF link with demodulate and forward relaying. Considering the FSO link in Gamma-Gamma atmospheric turbulence with the effect of pointing error in moderate to strong regime and the Negative Exponential atmospheric turbulence in saturate regime, and the RF link in Rayleigh fading, new closed form exact and asymptotic expressions are derived for the Outage Probability and Bit Error Rate of the proposed structure. Derived expressions are verified with MATLAB simulations. The results show that the performance of the proposed structure does not change with changing channel conditions; therefore, this structure is reliable and does not require more processing or power consumption in order to maintain system performance in poor channel conditions. Hence, the proposed structure is particularly suitable for mobile communications with limitations in power, processing and latency. The innovations and contributions of this paper, which are first introduced in the multi-hop hybrid FSO / RF hybrid structure, include the novelty of the proposed structure, the presentation of new exact and asymptotic mathematical solutions, the use of receive diversity, Bit Error Rate analysis, using amplify and forward as well as decode and forward protocols, taking into account a wide range of atmospheric turbulences with the effect of pointing error, using opportunistic selection and selection combining schemes.

## 1. Introduction

In Free Space Optical Communication (FSO) communication systems, often Intensity Modulation / Direct Detection (IM / DD) based on on-off keying (OOK) is used due to its simplicity [1]. The detection threshold of OOK adopts based on atmospheric turbulence intensity; therefore, it is suitable for variable atmospheric turbulence conditions. Pulse Position Modulation (PPM) and Subcarrier Intensity Modulation (SIM) schemes do not require adaptive detection threshold; spectral efficiency of SIM is more than PPM [2].

Differential modulations such as Differential Phase Shift Keying (DPSK) are less sensitive to noise and interference, and have optimal detection for the following reasons: no need for Channel State Information (CSI) or heavy computations at the receiver, no need for feedback to adjust the detection threshold, no effect on system throughput due to the lack of pilot or training sequence, reducing the effect of weather conditions such as fog and mist, reducing the effect of pointing error, reducing the effect of background noise at the receiver [3].

The FSO system has attracted the market for its numerous benefits, such as non-interference, lack of licensing or bandwidth management. It has a higher data rate and bandwidth than Radio Frequency (RF) system, and its installation is easy, fast and secure [4]. However, in spite of these advantages, practical implementation of FSO system is limited due to its high sensitivity to the effects of physical propagation media, such as atmospheric turbulence and pointing error.

The atmospheric turbulence causes the random oscillation of the received signal intensity [5]. Many statistical models have been presented to investigate the effects of atmospheric turbulence including Log-Normal [6], Gamma-Gamma [7], K [8], M [9], and Negative Exponential [10]. Among them Gamma-Gamma and Negative Exponential are in high accompany with experimental results for moderate to strong, and saturate regimes, respectively [11]. The M and K distributions provide excellent matching and agreement between theoretical and experimental data. The M distribution is capable of characterizing most of the existing models, including the Gamma-Gamma and the K distribution [8]. The K distribution is a suitable model for strong turbulence conditions [12].

The presence of wind, weak earthquake, and building vibration cause misalignment of the FSO transmitter and receiver [13]. This effect is called pointing error and severely degrades the performance of the FSO system [14].

The FSO and RF links are complementary of each other; there is almost no condition that could disrupt both of them at the same time. Therefore, combining FSO and RF links is a good idea to achieve advantages such as high reliability, accessibility, and data rate. The so-called hybrid FSO / RF systems are available in the series [15] or parallel [7] structure. In the series structure, the transmitted data is received, processed, and forwarded by a relay, and each link can be either RF or FSO [16]. In the parallel structure, an RF link is paralleled to the FSO link, and data is transmitted simultaneously [17] or by using a switch [18].

In the past decade, various works have been done in conjunction with the relay-assisted FSO system. The results of these researches indicate the efficiency and performance improve of this system [19]. The processing protocols of relay-assisted FSO system include detect and forward [20], decode and forward [19], and amplify and forward [21]. Among them, amplify and forward is the best choice for long range mobile communication, because long range communication requires consuming a lot of power at the transmitter side; but the mobile transmitter cannot provide much power. Therefore, it is better to use a relay as a signal amplifier. When communication is more important than power consumption, it is better to use a fixed gain amplify and forward because in this scheme the amplification gain is adjusted manually based on the worst case scenario; therefore, the amplified signal will show favorable performance in all situations.

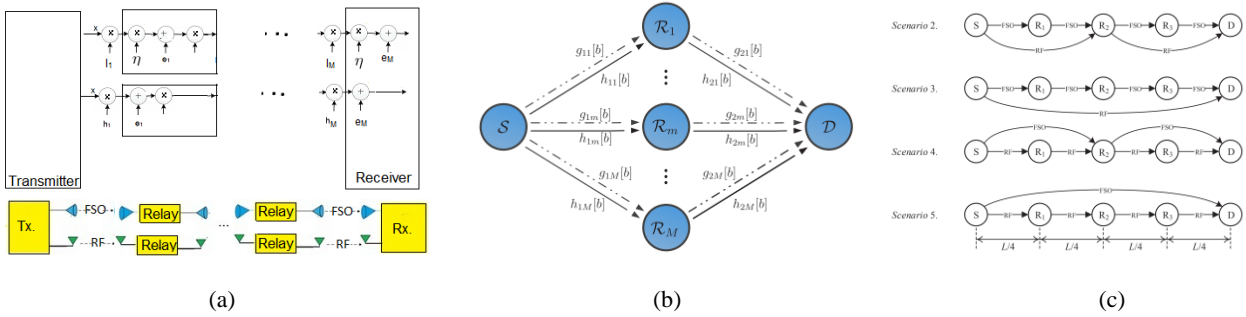

| (a) | (b) | (c) |

**Fig. 1:** (a) the proposed structure in Ref. [44,45], (b) the proposed structure in Ref. [42], (c) the proposed structure in Ref. [40].

However, amplify and forward, due to high power consumption, is not so cost-effective; it is only recommended in long range links or conditions that connection implementation is more important than cost or power consumption. Therefore, in other conditions, such as multi-hop links that have short links between hops, this scheme is not recommended and the use of other relay protocols, such as demodulate and forward as well as decode and forward, are preferable. It's worth noting that the complexity of demodulate and forward is less than decode and forward.

The proposed multi-hop hybrid FSO / RF structure of this paper is consisted of two main parts. The first part establishes the communication of a mobile user with the source base station at a long range link. At this part, the transmitted RF signal from the mobile user is amplified by a fixed gain relay and forwarded to the source base station through a hybrid parallel FSO / RF link. To have better performance, the relay uses multiple receive antennas with selection combining scheme, and the data transmission at hybrid parallel FSO / RF link is simultaneously. The second part establishes the communication of the source and the destination base stations via a multi-hop hybrid parallel FSO / RF system with demodulate and forward relaying. At this part, in order to have better performance, between received FSO and RF signals at the source base station, the signal with higher SNR is demodulated, regenerated, then modulated and forwarded through a hybrid parallel FSO / RF link simultaneously. This procedure repeats in all of the subsequent relays until reaching the destination base station.

The FSO link has Gamma-Gamma atmospheric turbulence with the effect of pointing error in moderate to strong regimes, and Negative Exponential atmospheric turbulence in saturate regime; RF link has Rayleigh fading. Though these channel models are frequently used in literatures, it should be noted that the aim of this paper is not to explore a new channel model; it is to present a new structure and prove that the proposed structure has worth to be practically implemented; these channel models are sufficient to this end. New exact and asymptotic expressions are derived in closed form for Bit Error Rate (BER) and Outage Probability of the proposed structure. Derived expressions are verified by MATLAB simulations. At the first part of the proposed structure, a fixed gain amplify and forward relay with multiple receive antennas amplifies the mobile transmitted signal. Therefore, this structure is particularly recommended for mobile communications limitations in power, processing, and latency. At the second part of the proposed structure hybrid multi-hop parallel FSO / RF simultaneous data transmission plus opportunistic signal selection significantly improves performance and capacity of the system.

The remainder of this paper is organized as follows: section 2 reviews some of the published works in hybrid FSO / RF system; in sections 3, 4, and 5, system model, Outage Probability, and BER of the proposed structure are discussed, respectively. Section 6 compares analytical and simulation results, and section 7 is conclusion of this study.

## 2. Related Works

The works done on the hybrid FSO / RF system divide into three main categories. The first category is the single-hop hybrid FSO / RF structure [23-29]. Few works of this category investigated diversity schemes [25]. The second group examines the performance of the dual-hop structure [2, 12, 13, 15, 30-39]; in this area, few works used diversity schemes [39]; also parallel transmission was investigated in few works [31]. The third category involves the multi-hop structure [40-45]. The multi-hop structure was previously investigated in FSO systems [46-50]; but in FSO / RF system, it is known as a new issue in emerged in recent years. The existing multi-hop FSO / RF structures investigated Outage Probability of a multi-hop hybrid FSO / RF link [40, 42, 44, 45].

According to the best of authors' knowledge, before online pre-printing of this work, there were published only 4 works on multi-hop hybrid FSO / RF system [40, 42, 44, 45]; all of them have investigated Outage Probability, but BER remained as an Open Problem. Because, in order to calculate the BER of a hybrid multi-hop FSO / RF system should integrate the multiplication of at least three Meijer-G functions and one exponential function (see Eq. (31)). To the best of authors' knowledge, there is no exact or asymptotic solution for such an integration. The ideas of this paper for exact and asymptotic solutions are taken from [51] and [12], respectively. Though [51, 12] brought the idea to the mind, they investigated single-hop FSO structures.

In order to have a better comparison, the existing multi-hop structures are reviewed in summary. Fig. 1 (a) displays the proposed structure of Ref. [44, 45]; a multi-hop single-receive antenna series FSO / RF system. The FSO link between relay and destination has Gamma-Gamma and Exponential atmospheric turbulence; the RF link has Rayleigh & Rican fading. The first hop is FSO / RF. Fig. 1 (b) displays the proposed structure in Ref. [42]; a multi-hop single-receive antenna parallel FSO / RF system. FSO link between relay and destination has Gamma-Gamma atmospheric turbulence; the RF link has Rican fading. First hop is parallel multi-hop. The effect of pointing error is not considered. Fig. 1 (c) displays the proposed structure in Ref. [40]; a multi-hop single-receive antenna series FSO / RF system. FSO link between relay and destination has log-normal atmospheric turbulence; RF link has Rican fading. First hops are different FSO / RF. The effect of pointing error is not considered.

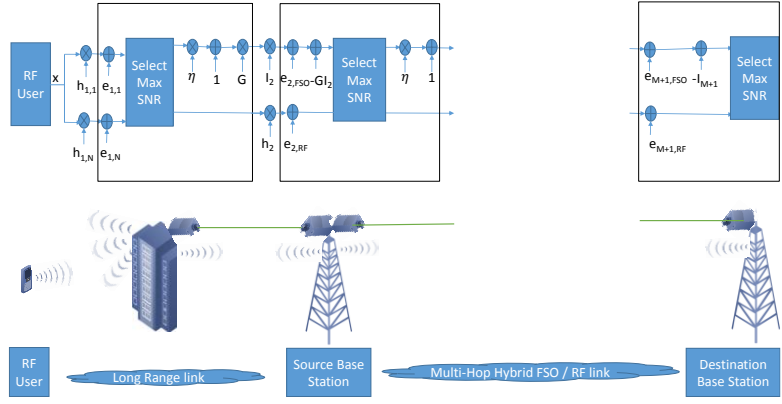

**Fig. 2.** The proposed multi-hop hybrid FSO / RF system.

Fig. 2 displays the proposed structure of this paper; a multi-hop multi-receive antenna series-parallel hybrid FSO / RF system with signal selection at each relay. Demodulate and forward relaying plus amplify and forward relaying are used. FSO link has Gamma-Gamma and Negative Exponential atmospheric turbulences; RF link has Rayleigh fading. The effect of pointing error is considered. First hop is RF. Outage Probability and BER are investigated. New closed form exact and asymptotic expressions are derived in for BER and Outage Probability.

According to the best of the authors' knowledge, the main contributions and innovations of this paper, which are introduced in multi-hop hybrid FSO / RF system for the first time, include the following: presenting a new structure, providing a new mathematical solution to the problem of system performance evaluation, the use of the receive diversity scheme, the use of selection combining and opportunistic selection schemes, the use of amplify and forward as well as demodulate and forward relaying, considering a wide range of atmospheric turbulences from medium to saturate regimes with the effect of pointing error, deriving new closed form exact and asymptotic expressions for Outage Probability and BER.

## 3. System Model

The proposed multi-hop hybrid FSO / RF structure is presented in Fig. 2. This structure includes two main parts. At the first part of this structure, a fixed gain amplify and forward relay with multiple receive antennas connects the mobile user to the source base station through a dual-hop hybrid FSO / RF system. At the second part of this structure, source and destination base stations are connected via a multi-hop hybrid parallel FSO / RF link with demodulate and forward relying.

Consider Fig. 2, let $x$ be the transmitted signal from mobile user, the received signal at $i-th$ receiver antenna of the first relay becomes as follows:

$$y_{1,i} = h_{1,i}x + e_{1,i}, \quad (1)$$

where $e_{1,i}$ is Additive White Gaussian Noise (AWGN), with zero mean and $\sigma_{RF}^2$ variance, at the input of $i-th$ receive antenna of the first relay, and $h_{1,i}$ is the fading coefficient of $i-th$ path between mobile user and first relay. The first relay uses selection combining scheme, so between received RF signals, the signal with the highest Signal to Noise Ratio (SNR) is selected; then the selected signal is duplicated. One copy of the duplicated signal is converted to FSO signal by conversion efficiency of $\eta$ and summed up with a bias, because the FSO signal is not negative. Then the FSO signal and the other copy of the duplicated signal are amplified with fixed gain of $G$ and simultaneously forwarded through parallel hybrid FSO / RF link. The forwarded FSO and RF signals are respectively as follows:

$$x_{2,FSO} = G(1 + \eta y_1), \quad (2)$$
$$x_{2,RF} = Gy_1, \quad (3)$$

where $y_1$ is the selected signal at the first relay. After DC removal, the received FSO and RF signals at the second relay (source Base Station) become respectively as follows:

$$y_{2,FSO} = G\eta I_2 h_1 x + G\eta I_2 e_1 + e_{2,FSO}, \quad (4)$$
$$y_{2,RF} = Gh_2 h_1 x + Gh_2 e_1 + e_{2,RF}, \quad (5)$$

where $I_2$ and $h_2$ are respectively the atmospheric turbulence intensity and the fading coefficient of the second hop, $e_{2,RF}$ and $e_{2,FSO}$ are AWGN with zero mean, $\sigma_{RF}^2$ and $\sigma_{FSO}^2$ variance, at the RF and FSO receiver inputs, respectively. Accordingly, the instantaneous SNR at the FSO and RF receiver inputs of the second relay become respectively as follows:

$$\gamma_{2^{nd}relay,FSO} = \frac{G^2\eta^2 I_2^2 h_1^2}{G^2\eta^2 I_2^2 \sigma_{RF}^2 + \sigma_{FSO}^2}, \quad (6)$$

$$\gamma_{2^{nd}relay,RF} = \frac{G^2 h_2^2 h_1^2}{G^2 h_2^2 \sigma_{RF}^2 + \sigma_{RF}^2}. \quad (7)$$

Regarding the fixed gain amplification, $G$ could be chosen as $G^2 = 1/(C\sigma_{RF}^2)$, where $C$ is a desired constant parameter [12]. Substituting $\gamma_1 = \frac{h_1^2}{\sigma_{RF}^2}$, $\gamma_{2,RF} = \frac{G^2 h_2^2}{\sigma_{RF}^2}$ and $\gamma_{2,FSO} = \frac{G^2\eta^2 I_2^2}{\sigma_{FSO}^2}$, and $G$, (6) and (7) become as follows:

$$\gamma_{2^{nd}relay,FSO} = \frac{\gamma_1 \gamma_{2,FSO}}{C + \gamma_{2,FSO}}, \quad (8)$$

$$\gamma_{2^{nd}relay,RF} = \frac{\gamma_1 \gamma_{2,RF}}{C + \gamma_{2,RF}}. \quad (9)$$

In order to have better performance, between received FSO and RF signals at the second relay, one with higher SNR is selected.

Then the selected signal is demodulated, regenerated, and duplicated. The same as the first relay, one copy is transformed to FSO signal; then the FSO signal and the other copy of the duplicated signal are modulated and forwarded simultaneously through hybrid parallel FSO / RF link. The procedure described in this paragraph repeats in all subsequent relays until reaching the destination base station.

The probability density function (pdf) and Cumulative Distribution Function (CDF) of Gamma-Gamma atmospheric turbulence with the effect of pointing error [51], CDF of Rayleigh fading and Negative Exponential atmospheric turbulence, are respectively as follows:

$$f_\gamma(\gamma) = \frac{\xi^2}{2\Gamma(\alpha)\Gamma(\beta)\gamma} G_{1,3}^{3,0}\left(\alpha\beta\kappa\sqrt{\frac{\gamma}{\bar{\gamma}_{FSO}}} \Big| \begin{matrix} \xi^2 + 1 \\ \xi^2, \alpha, \beta \end{matrix}\right), \quad (10)$$

$$F_\gamma(\gamma) = \frac{\xi^2}{\Gamma(\alpha)\Gamma(\beta)} G_{2,4}^{3,1}\left(\alpha\beta\kappa\sqrt{\frac{\gamma}{\bar{\gamma}_{FSO}}} \Big| \begin{matrix} 1, \xi^2 + 1 \\ \xi^2, \alpha, \beta, 0 \end{matrix}\right), \quad (11)$$

$$F_\gamma(\gamma) = 1 - e^{-\frac{\gamma}{\bar{\gamma}_{RF}}}, \quad (12)$$

$$F_\gamma(\gamma) = 1 - e^{-\lambda\sqrt{\frac{\gamma}{\bar{\gamma}_{FSO}}}}, \quad (13)$$

where $G_{p,q}^{m,n}\left(z \Big| \begin{matrix} a_1, a_2, \dots, a_p \\ b_1, b_2, \dots, b_q \end{matrix}\right)$ is Meijer-G function [52, Eq. 07.34.02.0001.01], $\Gamma(.)$ is Gamma function [52, Eq. 06.05.02.0001.01], $\xi^2 = \omega_{Z_{eq}}/(2\sigma_s)$ is the ratio of the equivalent received beam radius $(\omega_{Z_{eq}}/2)$ to the standard deviation of pointing errors at the receiver $(\sigma_s)$, $\kappa = \frac{\xi^2}{\xi^2+1}$, $\alpha = \left[\exp\left(0.49\sigma_R^2/(1+1.11\sigma_R^{12/5})^{7/6}\right) - 1\right]^{-1}$ and $\beta = \left[\exp\left(0.51\sigma_R^2/(1+0.69\sigma_R^{12/5})^{5/6}\right) - 1\right]^{-1}$ are parameters related to Gamma-Gamma atmospheric turbulence, where $\sigma_R^2$ is Rytov variance [36]. $\bar{\gamma}_{FSO} = \eta^2/\sigma_{FSO}^2$ and $\bar{\gamma}_{RF} = 1/\sigma_{RF}^2$ are average SNR at the FSO and RF receiver input, and $1/\lambda^2$ is variance of Negative Exponential distribution.

The receiving antennas of the first relay are assumed to be separated from each other; therefore, the received signal by each of them experiences independent fading. Thus, with high probability, it could be said that at any given time slot, at least one signal experiences desired condition. Accordingly, if always use best signal, the overall system performance improves. In this paper, the signal with the best link quality is considered as the best signal. Accordingly, the CDF of the instantaneous SNR at the first input of the relay ($\gamma_1$) becomes as follows:

$$F_{\gamma_1}(\gamma) = Pr(max(\gamma_{1,1}, \gamma_{1,2}, \dots, \gamma_{1,N}) \leq \gamma) =$$
$$Pr(\gamma_{1,1} \leq \gamma, \gamma_{1,2} \leq \gamma, \dots, \gamma_{1,N} \leq \gamma) = \prod_{i=1}^{N} Pr(\gamma_{1,i} \leq \gamma) = \left(F_{\gamma_{1,i}}(\gamma)\right)^N = \left(1 - e^{-\frac{\gamma}{\bar{\gamma}_{RF}}}\right)^N, \quad (14)$$

By derivation of (14) and using the binomial expansion theorem, the pdf of $\gamma_1$ becomes as follows:

$$f_{\gamma_1}(\gamma) = \frac{N}{\bar{\gamma}_{RF}} \sum_{k=0}^{N-1} \binom{N-1}{k} (-1)^k e^{-\frac{(k+1)\gamma}{\bar{\gamma}_{RF}}}. \quad (15)$$

In order to improve the performance, at the $j - th$; $j = 2,4, \dots M$ relay, between received FSO and RF signals, one with higher SNR is selected. Therefore, the CDF of the instantaneous SNR at the $j - th$ relay input ($\gamma_j$) becomes as follows:

$$F_{\gamma_j}(\gamma) = Pr(max(\gamma_{FSO,j}, \gamma_{RF,j}) \leq \gamma) = Pr(\gamma_{FSO,j} \leq \gamma, \gamma_{RF,j} \leq \gamma) = F_{\gamma_{FSO,j}}(\gamma) F_{\gamma_{RF,j}}(\gamma). \quad (16)$$

The last equality appears because of independence of FSO and RF links. According to (8) and (9), the CDFs of $\gamma_{2^{nd}relay,FSO}$ and $\gamma_{2^{nd}relay,RF}$ become as follows [36]:

$$F_{\gamma_{2^{nd}relay,FSO}}(\gamma) = 1 - \int_0^\infty Pr\left(\gamma_{2,FSO} \geq \frac{\gamma C}{x} \Big| \gamma_1\right) f_{\gamma_1}(x+\gamma) dx, \quad (17)$$

$$F_{\gamma_{2^{nd}relay,RF}}(\gamma) = 1 - \int_0^\infty Pr\left(\gamma_{2,RF} \geq \frac{\gamma C}{x} \Big| \gamma_1\right) f_{\gamma_1}(x+\gamma) dx. \quad (18)$$

By substituting (12) and (15) into (17) and using [52, Eq.07.34.17.0012.01], the CDF of $\gamma_{2^{nd}relay,FSO}$ in Gamma-Gamma atmospheric turbulence with the effect of pointing error becomes equal to:

$$F_{\gamma_{2^{nd}relay,FSO}}(\gamma) = 1 - $$
$$\sum_{k=0}^{N-1} \binom{N-1}{k} (-1)^k \frac{N}{\bar{\gamma}_{RF}} e^{-\frac{(k+1)\gamma}{\bar{\gamma}_{RF}}} \int_0^\infty e^{-\frac{(k+1)x}{\bar{\gamma}_{RF}}} \left(1 - \frac{\xi^2}{\Gamma(\alpha)\Gamma(\beta)} G_{4,2}^{1,3}\left(\frac{1}{\alpha\beta\kappa}\sqrt{\frac{x\bar{\gamma}_{FSO}}{\gamma C}} \Big| \begin{matrix} 1-\xi^2, 1-\alpha, 1-\beta, 1 \\ 0, -\xi^2 \end{matrix}\right)\right) dx. \quad (19)$$

By using [52, Eq. 07.34.21.0088.01] and [53, Eq. 07.34.17.0012.01], the CDF of $\gamma_{2^{nd}relay,FSO}$ in Gamma-Gamma atmospheric turbulence with the effect of pointing error becomes equal to:

$$F_{\gamma_{2^{nd}relay,FSO}}(\gamma) = 1 - \sum_{k=0}^{N-1} \binom{N-1}{k} \frac{(-1)^k N}{k+1} e^{-\frac{(k+1)\gamma}{\bar{\gamma}_{RF}}} \left(1 - \frac{\xi^2 2^{\alpha+\beta-3}}{\pi\Gamma(\alpha)\Gamma(\beta)} G_{4,9}^{7,2}\left(\frac{(\alpha\beta\kappa)^2 C(k+1)\gamma}{16\bar{\gamma}_{FSO}\bar{\gamma}_{RF}} \Big| \begin{matrix} \psi_1 \\ \psi_2 \end{matrix}\right)\right), \quad (20)$$

where $\psi_1 = \left\{1, \frac{1}{2}, \frac{\xi^2+2}{2}, \frac{\xi^2+1}{2}\right\}$ and $\psi_2 = \left\{1, \frac{1}{2}, \frac{\xi^2+1}{2}, \frac{\xi^2}{2}, \frac{\alpha+1}{2}, \frac{\alpha}{2}, \frac{\beta+1}{2}, \frac{\beta}{2}, \frac{1}{2}, 0\right\}$.

By substituting (13) and (15) into (17) and substituting equivalent Meijer-G function of $e^{-\lambda\sqrt{\frac{C\gamma}{x\bar{\gamma}_{FSO}}}}$ as $\frac{1}{\sqrt{\pi}} G_{2,0}^{0,2}\left(\frac{4x\bar{\gamma}_{FSO}}{\lambda^2\gamma C} \Big| \begin{matrix} 1,1/2 \\ - \end{matrix}\right)$ using [52, Eq.07.34.03.1081.01] and [52, Eq. 07.34.17.0012.01], the CDF of $\gamma_{2^{nd}relay,FSO}$ in Negative Exponential atmospheric turbulence becomes equal to:

$$F_{\gamma_{2^{nd}relay,FSO}}(\gamma) = 1 - \sum_{k=0}^{N-1} \binom{N-1}{k} \frac{(-1)^k N}{\bar{\gamma}_{RF}} e^{-\frac{(k+1)\gamma}{\bar{\gamma}_{RF}}} \times$$
$$\int_0^\infty e^{-\frac{(k+1)x}{\bar{\gamma}_{RF}}} \frac{1}{\sqrt{\pi}} G_{2,0}^{0,2}\left(\frac{4x\bar{\gamma}_{FSO}}{\lambda^2\gamma C} \Big| \begin{matrix} 1,1 \\ 2 \end{matrix}\right) dx. \quad (21)$$

By using [52, Eq. 07.34.17.0012.01] and [52, Eq. 07.34.21.0088.01], the CDF of $\gamma_{2^{nd}relay,FSO}$ in Negative Exponential atmospheric turbulence becomes equal to:

$$F_{\gamma_{2^{nd}relay,FSO}}(\gamma) = 1 - $$
$$\sum_{k=0}^{N-1} \binom{N-1}{k} \frac{(-1)^k N}{\sqrt{\pi}(k+1)} e^{-\frac{(k+1)\gamma}{\bar{\gamma}_{RF}}} G_{0,3}^{3,0}\left(\frac{\lambda^2 C(k+1)\gamma}{4\bar{\gamma}_{FSO}\bar{\gamma}_{RF}} \Big| \begin{matrix} - \\ 1,0,\frac{1}{2} \end{matrix}\right). \quad (22)$$

By substituting (14) and (15) into (18), and substituting equivalent Meijer-G function of $e^{-\frac{\gamma C}{x\bar{\gamma}_{RF}}}$ as $G_{1,0}^{0,1}\left(\frac{x\bar{\gamma}_{RF}}{\gamma C} \Big| \begin{matrix} 1 \\ - \end{matrix}\right)$ using [52, Eq. 07.34.03.1081.01] and [52, Eq.07.34.17.0012.01], the CDF of $\gamma_{2^{nd}relay,RF}$ in Rayleigh fading becomes equal to:

$$F_{\gamma_{2^{nd}relay,RF}}(\gamma) = 1 - $$
$$\sum_{k=0}^{N-1} \binom{N-1}{k} \frac{(-1)^k N}{\bar{\gamma}_{RF}} e^{-\frac{(k+1)\gamma}{\bar{\gamma}_{RF}}} \int_0^\infty e^{-\frac{(k+1)x}{\bar{\gamma}_{RF}}} G_{1,0}^{0,1}\left(\frac{x\bar{\gamma}_{RF}}{\gamma C} \Big| \begin{matrix} 1 \\ - \end{matrix}\right) dx. \quad (23)$$

Using [52, Eq.07.34.21.0088.01] and [52,Eq.07.34.17.0012.01], the CDF of $\gamma_{2^{nd}relay,RF}$ in Rayleigh fading becomes equal to:

$$F_{\gamma_{2^{nd}relay,RF}}(\gamma) = 1 - \sum_{k=0}^{N-1}\binom{N-1}{k}\frac{(-1)^k N}{k+1}e^{-\frac{(k+1)\gamma}{\bar{\gamma}_{RF}}}G_{0,2}^{2,0}\left(\frac{C(k+1)\gamma}{\bar{\gamma}_{RF}^2}\Big|_{1,0}^{-}\right). \quad (24)$$

## 4. Outage Probability

In the proposed structure, the outage occurs when the instantaneous SNR of each relay becomes lower than a threshold. By assuming the independent operation of each relay, the availability probability of the proposed structure equals to the multiplication of the availability probabilities of individual relays [40]; considering this fact and the independence of the FSO and RF links, Outage Probability of the proposed structure is as follows:

$$P_{out}(\gamma_{th}) = Pr\{(\gamma_1,\gamma_2,\ldots,\gamma_{M+1}) \leq \gamma_{th}\} = 1 - Pr\{\gamma_1 \geq \gamma_{th}, \gamma_2 \geq \gamma_{th}, \ldots, \gamma_{M+1} \geq \gamma_{th}\} = 1 - Pr\{\gamma_1 \geq \gamma_{th}\}Pr\{\gamma_2 \geq \gamma_{th}\}\ldots Pr\{\gamma_{M+1} \geq \gamma_{th}\} = 1 - (1 - Pr\{(\gamma_1,\gamma_2)\leq \gamma_{th}\})(1 - Pr\{(\gamma_3 \leq \gamma_{th})\})(1 - Pr\{(\gamma_{M+1} \leq \gamma_{th})\}) = 1 - \left(1 - P_{out,2^{nd}relay}(\gamma_{th})\right)\left(1 - P_{out,3}(\gamma_{th})\right)\ldots\left(1 - P_{out,M+1}(\gamma_{th})\right) = 1 - \left(1 - F_{\gamma_{2^{nd}relay,FSO}}(\gamma_{th})F_{\gamma_{2^{nd}relay,RF}}(\gamma_{th})\right)\left(1 - F_{\gamma_{j,FSO}}(\gamma_{th})F_{\gamma_{j,RF}}(\gamma_{th})\right)^{M-1}.$$

(25)

### 4.1 Gamma-Gamma atmospheric turbulence with the effect of pointing error

By substituting (12), (11), (20) and (24) into (25), Outage Probability of the proposed structure in Gamma-Gamma atmospheric turbulence with the effect of pointing error becomes as (26).

By using binomial expansion of $\left[1 - \frac{\xi^2}{\Gamma(\alpha)\Gamma(\beta)}\left(1 - e^{-\frac{\gamma_{th}}{\bar{\gamma}_{RF}}}\right) \times G_{2,4}^{3,1}\left(\alpha\beta\kappa\sqrt{\frac{\gamma_{th}}{\bar{\gamma}_{FSO}}}\Big|_{\xi^2,\alpha,\beta,0}^{1,\xi^2+1}\right)\right]^{M-1}$ as $\sum_{t=0}^{M-1}\sum_{u=0}^{t}\binom{M-1}{t}\binom{t}{u}(-1)^{t+u} \times e^{-\frac{u\gamma_{th}}{\bar{\gamma}_{RF}}}\left(\frac{\xi^2}{\Gamma(\alpha)\Gamma(\beta)}G_{2,4}^{3,1}\left(\alpha\beta\kappa\sqrt{\frac{\gamma_{th}}{\bar{\gamma}_{FSO}}}\Big|_{\xi^2,\alpha,\beta,0}^{1,\xi^2+1}\right)\right)^t$, Outage Probability of the proposed structure in Gamma-Gamma atmospheric turbulence with the effect of pointing error becomes as (27), where $\Omega = \binom{N-1}{k}\binom{M-1}{t}\binom{t}{u}(-1)^{k+t+u}\frac{N}{k+1}$.

Equation (27) and other derived expressions (in the following) of this paper are complex; therefore, it is not easy to have insight about them. However, authors tried to provide physical insights at the results section. Complexity of derived expressions is related to the complexity of the proposed structure and the complexity of Meijer-G function. To the best of the authors' knowledge, this is the first time that such a complicated structure is investigated in hybrid FSO / RF system.

$$P_{out}(\gamma_{th}) = 1 - \left[\sum_{k=0}^{N-1}\binom{N-1}{k}\frac{(-1)^k N}{k+1}e^{-\frac{(k+1)\gamma_{th}}{\bar{\gamma}_{RF}}}\left(1 - \frac{\xi^2 2^{\alpha+\beta-3}}{\pi\Gamma(\alpha)\Gamma(\beta)}G_{4,9}^{7,2}\left(\frac{(\alpha\beta\kappa)^2 C(k+1)\gamma_{th}}{16\bar{\gamma}_{FSO}\bar{\gamma}_{RF}}\Big|_{\psi_2}^{\psi_1}\right)\right) + \sum_{k=0}^{N-1}\binom{N-1}{k}\times \frac{(-1)^k N}{k+1}e^{-\frac{(k+1)\gamma_{th}}{\bar{\gamma}_{RF}}}G_{0,2}^{2,0}\left(\frac{C(k+1)\gamma_{th}}{\bar{\gamma}_{RF}^2}\Big|_{1,0}^{-}\right) - \sum_{k=0}^{N-1}\sum_{g=0}^{N-1}\binom{N-1}{k}\binom{N-1}{g}\times\frac{(-1)^{g+k}N^2}{(k+1)(g+1)}e^{-\frac{(k+g+1)\gamma_{th}}{\bar{\gamma}_{RF}}}G_{0,2}^{2,0}\left(\frac{C(g+1)\gamma_{th}}{\bar{\gamma}_{RF}^2}\Big|_{1,0}^{-}\right)\left(1 - \frac{\xi^2 2^{\alpha+\beta-3}}{\pi\Gamma(\alpha)\Gamma(\beta)}G_{4,9}^{7,2}\left(\frac{(\alpha\beta\kappa)^2 C(k+1)\gamma_{th}}{16\bar{\gamma}_{FSO}\bar{\gamma}_{RF}}\Big|_{\psi_2}^{\psi_1}\right)\right)\right]\left[1 - \frac{\xi^2}{\Gamma(\alpha)\Gamma(\beta)}\left(1 - e^{-\frac{\gamma_{th}}{\bar{\gamma}_{RF}}}\right)G_{2,3}^{3,1}\left(\alpha\beta\kappa\sqrt{\frac{\gamma_{th}}{\bar{\gamma}_{FSO}}}\Big|_{\xi^2,\alpha,\beta,0}^{1,\xi^2+1}\right)\right]^{M-1} \quad (26)$$

$$P_{out}(\gamma_{th}) = 1 - \sum_{k=0}^{N-1}\sum_{t=0}^{M-1}\sum_{u=0}^{t}\Omega\, e^{-\frac{(k+u+1)\gamma_{th}}{\bar{\gamma}_{RF}}}\left(\frac{\xi^2}{\Gamma(\alpha)\Gamma(\beta)}G_{2,4}^{3,1}\left(\alpha\beta\kappa\sqrt{\frac{\gamma_{th}}{\bar{\gamma}_{FSO}}}\Big|_{\xi^2,\alpha,\beta,0}^{1,\xi^2+1}\right)\right)^t\left(1 - \frac{\xi^2 2^{\alpha+\beta-3}}{\pi\Gamma(\alpha)\Gamma(\beta)}\times G_{4,9}^{7,2}\left(\frac{(\alpha\beta\kappa)^2 C(k+1)\gamma_{th}}{16\bar{\gamma}_{FSO}\bar{\gamma}_{RF}}\Big|_{\psi_2}^{\psi_1}\right)\right) - \sum_{k=0}^{N-1}\sum_{t=0}^{M-1}\sum_{u=0}^{t}\Omega\, e^{-\frac{(k+u+1)\gamma_{th}}{\bar{\gamma}_{RF}}}\left(\frac{\xi^2}{\Gamma(\alpha)\Gamma(\beta)}G_{2,4}^{3,1}\left(\alpha\beta\kappa\sqrt{\frac{\gamma_{th}}{\bar{\gamma}_{FSO}}}\Big|_{\xi^2,\alpha,\beta,0}^{1,\xi^2+1}\right)\right)^t \times G_{0,2}^{2,0}\left(\frac{\gamma_{th}C(k+1)}{\bar{\gamma}_{RF}^2}\Big|_{1,0}^{-}\right) + \sum_{k=0}^{N-1}\sum_{g=0}^{N-1}\sum_{t=0}^{M-1}\sum_{u=0}^{t}\binom{N-1}{g}(-1)^g\frac{N\Omega}{g+1}e^{-\frac{(k+g+u+2)\gamma_{th}}{\bar{\gamma}_{RF}}}\left(\frac{\xi^2}{\Gamma(\alpha)\Gamma(\beta)}\times G_{2,4}^{3,1}\left(\alpha\beta\kappa\sqrt{\frac{\gamma_{th}}{\bar{\gamma}_{FSO}}}\Big|_{\xi^2,\alpha,\beta,0}^{1,\xi^2+1}\right)\right)^t G_{0,2}^{2,0}\left(\frac{\gamma_{th}C(k+1)}{\bar{\gamma}_{RF}^2}\Big|_{1,0}^{-}\right)\left(1 - \frac{\xi^2 2^{\alpha+\beta-3}}{\pi\Gamma(\alpha)\Gamma(\beta)}G_{4,9}^{7,2}\left(\frac{(\alpha\beta\kappa)^2 C(g+1)\gamma_{th}}{16\bar{\gamma}_{FSO}\bar{\gamma}_{RF}}\Big|_{\psi_2}^{\psi_1}\right)\right) \quad (27)$$

$$P_{out}(\gamma_{th}) = 1 - \left[\sum_{k=0}^{N-1}\binom{N-1}{k}\frac{(-1)^k N}{\sqrt{\pi}(k+1)}e^{-\frac{(k+1)\gamma_{th}}{\bar{\gamma}_{RF}}}G_{0,3}^{3,0}\left(\frac{\lambda^2 C(k+1)\gamma_{th}}{4\bar{\gamma}_{FSO}\bar{\gamma}_{RF}}\Big|_{1,0,\frac{1}{2}}^{-}\right) + \sum_{k=0}^{N-1}\binom{N-1}{k}\frac{(-1)^k N}{k+1}e^{-\frac{(k+1)\gamma_{th}}{\bar{\gamma}_{RF}}}G_{0,2}^{2,0}\left(\frac{C(k+1)\gamma_{th}}{\bar{\gamma}_{RF}^2}\Big|_{1,0}^{-}\right) - \sum_{k=0}^{N-1}\sum_{g=0}^{N-1}\binom{N-1}{k}\binom{N-1}{g}\frac{(-1)^{g+k}N^2}{\sqrt{\pi}(k+1)(g+1)}e^{-\frac{(k+g+1)\gamma_{th}}{\bar{\gamma}_{RF}}}G_{0,3}^{3,0}\left(\frac{\lambda^2 C(g+1)\gamma_{th}}{4\bar{\gamma}_{FSO}\bar{\gamma}_{RF}}\Big|_{1,0,\frac{1}{2}}^{-}\right)\left(1 - \frac{\xi^2 2^{\alpha+\beta-3}}{\pi\Gamma(\alpha)\Gamma(\beta)}G_{4,9}^{7,2}\left(\frac{(\alpha\beta\kappa)^2 C(k+1)\gamma_{th}}{16\bar{\gamma}_{FSO}\bar{\gamma}_{RF}}\Big|_{\psi_2}^{\psi_1}\right)\right)\right]\left[1 - \frac{\xi^2}{\Gamma(\alpha)\Gamma(\beta)}\left(1 - e^{-\frac{\gamma_{th}}{\bar{\gamma}_{RF}}}\right)G_{2,3}^{3,1}\left(\alpha\beta\kappa\sqrt{\frac{\gamma_{th}}{\bar{\gamma}_{FSO}}}\Big|_{\xi^2,\alpha,\beta,0}^{1,\xi^2+1}\right)\right]^{M-1} \quad (28)$$

$$P_{out}(\gamma_{th}) = 1 - \sum_{k=0}^{N-1}\sum_{t=0}^{M-1}\sum_{u=0}^{t}\sum_{v=0}^{t}\Lambda e^{-\frac{(k+u+1)\gamma_{th}}{\bar{\gamma}_{RF}}}e^{-\lambda v\sqrt{\frac{\gamma_{th}}{\bar{\gamma}_{FSO}}}}G_{0,2}^{2,0}\left(\frac{C(k+1)\gamma_{th}}{\bar{\gamma}_{RF}^2}\Big|_{1,0}^{-}\right) - \sum_{k=0}^{N-1}\sum_{t=0}^{M-1}\sum_{u=0}^{t}\sum_{v=0}^{t}\frac{\Lambda}{\sqrt{\pi}}e^{-\frac{(k+u+1)\gamma_{th}}{\bar{\gamma}_{RF}}}e^{-\lambda v\sqrt{\frac{\gamma_{th}}{\bar{\gamma}_{FSO}}}}G_{0,3}^{3,0}\left(\frac{\lambda^2 C(k+1)\gamma_{th}}{4\bar{\gamma}_{FSO}\bar{\gamma}_{RF}}\Big|_{1,0,1/2}^{-}\right) + \sum_{k=0}^{N-1}\sum_{g=0}^{N-1}\sum_{t=0}^{M-1}\sum_{u=0}^{t}\sum_{v=0}^{t}\binom{N-1}{g}(-1)^g\frac{N\Lambda}{\sqrt{\pi}(g+1)}e^{-\frac{(k+g+u+2)\gamma_{th}}{\bar{\gamma}_{RF}}}e^{-\lambda v\sqrt{\frac{\gamma_{th}}{\bar{\gamma}_{FSO}}}}G_{0,2}^{2,0}\left(\frac{C(k+1)\gamma_{th}}{\bar{\gamma}_{RF}^2}\Big|_{1,0}^{-}\right)\left(\frac{\lambda^2 C(g+1)\gamma_{th}}{4\bar{\gamma}_{FSO}\bar{\gamma}_{RF}}\Big|_{1,0,1/2}^{-}\right) \quad (29)$$

## 4.2 Negative Exponential atmospheric turbulence

By substituting (13), (12), (22) and (24) into (25), Outage Probability of the proposed structure in Negative Exponential atmospheric turbulence becomes as (28).

By using the binomial expansion of $\left[1-\left(1-e^{-\frac{\gamma_{th}}{\overline{\gamma}_{RF}}}\right)\left(1-e^{-\lambda\sqrt{\frac{\gamma_{th}}{\overline{\gamma}_{FSO}}}}\right)\right]^{M-1}$ as $\sum_{t=0}^{M-1}\sum_{u=0}^{t}\sum_{v=0}^{t}\binom{M-1}{t}\binom{t}{u}\binom{t}{v}(-1)^{t+u+v} \times e^{-\frac{u\gamma_{th}}{\overline{\gamma}_{RF}}}e^{-\lambda v\sqrt{\frac{\gamma_{th}}{\overline{\gamma}_{FSO}}}}$, Outage Probability of the proposed structure in Negative Exponential atmospheric turbulence becomes as (29), where $\Lambda = \binom{N-1}{k}\binom{M-1}{t}\binom{t}{u}\binom{t}{v}(-1)^{k+t+u+v}\frac{N}{k+1}$.

Though it is difficult to have insight about (29), it is are plotted and enough insights are provided in results section.

## 5. Bit Error Rate

Phase recovery error degrades performance of system with coherent modulations; whereas differential modulations such as DPSK are less sensitive to it, and their detection is less complex. The DPSK is of the most frequently used modulation formats in hybrid FSO / RF system. The BER of DPSK modulation can be calculated from the following equation [32]:

$$P_e = \frac{1}{2}\int_0^\infty e^{-\gamma} F_\gamma(\gamma) d\gamma = \frac{1}{2}\int_0^\infty e^{-\gamma} P_{out}(\gamma) d\gamma \tag{30}$$

where the last equality appeared because $F_\gamma(\gamma) = P_{out}(\gamma)$.

### 5.1 Gamma-Gamma atmospheric turbulence with the effect of pointing error

Substituting (27) into (30), BER of DPSK modulation in Gamma-Gamma atmospheric turbulence with the effect of pointing errors becomes as (31). In the case of $t > 0$, for solving (31) must integrate the multiplication of at least the three Meijer-G functions and one exponential function. To the best of the authors' knowledge, the exact or asymptomatic solution for this integration has not yet been provided. The idea of this paper for solving this integral is to obtain a linear substitution for $(.)^t$ in (31). In fact, after this substitution, at all $t$ values, for solving (31), it is sufficient to

$$P_e = \frac{1}{2}\int_0^\infty e^{-\gamma}\left\{1 - \sum_{k=0}^{N-1}\sum_{t=0}^{M-1}\sum_{u=0}^{t}\Omega\, e^{-\frac{(k+u+1)\gamma}{\overline{\gamma}_{RF}}}\left(\frac{\xi^2}{\Gamma(\alpha)\Gamma(\beta)}G_{2,4}^{3,1}\left(\alpha\beta\kappa\sqrt{\frac{\gamma}{\overline{\gamma}_{FSO}}}\bigg|\begin{matrix}1,\xi^2+1\\\xi^2,\alpha,\beta,0\end{matrix}\right)\right)^t \left(1 - \frac{\xi^2 2^{\alpha+\beta-3}}{\pi\Gamma(\alpha)\Gamma(\beta)}\times\right.\right. \tag{31}$$

$$G_{4,9}^{7,2}\left(\frac{(\alpha\beta\kappa)^2 C(k+1)\gamma}{16\overline{\gamma}_{FSO}\overline{\gamma}_{RF}}\bigg|\begin{matrix}\psi_1\\\psi_2\end{matrix}\right)\right) - \sum_{k=0}^{N-1}\sum_{t=0}^{M-1}\sum_{u=0}^{t}\Omega\, e^{-\frac{(k+u+1)\gamma}{\overline{\gamma}_{RF}}}\left(\frac{\xi^2}{\Gamma(\alpha)\Gamma(\beta)}G_{2,4}^{3,1}\left(\alpha\beta\kappa\sqrt{\frac{\gamma}{\overline{\gamma}_{FSO}}}\bigg|\begin{matrix}1,\xi^2+1\\\xi^2,\alpha,\beta,0\end{matrix}\right)\right)^t \times$$

$$G_{0,2}^{2,0}\left(\frac{\gamma C(k+1)}{\overline{\gamma}_{RF}^2}\bigg|\begin{matrix}-\\1,0\end{matrix}\right) + \sum_{k=0}^{N-1}\sum_{g=0}^{N-1}\sum_{t=0}^{M-1}\sum_{u=0}^{t}\binom{N-1}{g}(-1)^g\frac{N\Omega}{g+1}e^{-\frac{(k+g+u+2)\gamma}{\overline{\gamma}_{RF}}}\left(\frac{\xi^2}{\Gamma(\alpha)\Gamma(\beta)}\times\right.$$

$$G_{2,4}^{3,1}\left(\alpha\beta\kappa\sqrt{\frac{\gamma}{\overline{\gamma}_{FSO}}}\bigg|\begin{matrix}1,\xi^2+1\\\xi^2,\alpha,\beta,0\end{matrix}\right)\right)^t G_{0,2}^{2,0}\left(\frac{\gamma C(k+1)}{\overline{\gamma}_{RF}^2}\bigg|\begin{matrix}-\\1,0\end{matrix}\right)\left(1 - \frac{\xi^2 2^{\alpha+\beta-3}}{\pi\Gamma(\alpha)\Gamma(\beta)}G_{4,9}^{7,2}\left(\frac{(\alpha\beta\kappa)^2 C(g+1)\gamma}{16\overline{\gamma}_{FSO}\overline{\gamma}_{RF}}\bigg|\begin{matrix}\psi_1\\\psi_2\end{matrix}\right)\right)\right\}d\gamma$$

$$P_{out}(\gamma_{th}) = 1 - \sum_{k=0}^{N-1}\sum_{t=0}^{M-1}\sum_{u=0}^{t}\sum_{k_1=0}^{t}\sum_{k_2=0}^{k_1}\sum_{n=0}^{\infty}\Omega\binom{t}{k_1}\binom{k_1}{k_2}X_0^{t-k_1}\left(Y_n^{(k_1-k_2)}*Z_n^{(k_2)}\right)\left(\frac{\gamma_{th}}{\overline{\gamma}_{FSO}}\right)^{\beth}e^{-\frac{(k+u+1)\gamma_{th}}{\overline{\gamma}_{RF}}}\times \tag{34}$$

$$\left[1 - \frac{\xi^2 2^{\alpha+\beta-3}}{\pi\Gamma(\alpha)\Gamma(\beta)}G_{4,9}^{7,2}\left(\frac{(\alpha\beta\kappa)^2 C(k+1)\gamma_{th}}{16\overline{\gamma}_{FSO}\overline{\gamma}_{RF}}\bigg|\begin{matrix}\psi_1\\\psi_2\end{matrix}\right)\right] - \sum_{k=0}^{N-1}\sum_{t=0}^{M-1}\sum_{u=0}^{t}\sum_{k_1=0}^{t}\sum_{k_2=0}^{k_1}\sum_{n=0}^{\infty}\Omega\binom{t}{k_1}\binom{k_1}{k_2}X_0^{t-k_1}\times$$

$$\left(Y_n^{(k_1-k_2)}*Z_n^{(k_2)}\right)\left(\frac{\gamma_{th}}{\overline{\gamma}_{FSO}}\right)^{\beth}e^{-\frac{(k+u+1)\gamma_{th}}{\overline{\gamma}_{RF}}}G_{0,2}^{2,0}\left(\frac{C(k+1)\gamma_{th}}{\overline{\gamma}_{RF}^2}\bigg|\begin{matrix}-\\1,0\end{matrix}\right) +$$

$$\sum_{k=0}^{N-1}\sum_{g=0}^{N-1}\sum_{t=0}^{M-1}\sum_{u=0}^{t}\sum_{k_1=0}^{t}\sum_{k_2=0}^{k_1}\sum_{n=0}^{\infty}\Omega\binom{N-1}{g}\binom{t}{k_1}\binom{k_1}{k_2}(-1)^g X_0^{t-k_1}\left(Y_n^{(k_1-k_2)}*\right.$$

$$\left.Z_n^{(k_2)}\right)\frac{N}{g+1}\left(\frac{\gamma_{th}}{\overline{\gamma}_{FSO}}\right)^{\beth}e^{-\frac{(k+g+u+2)\gamma_{th}}{\overline{\gamma}_{RF}}}\times G_{0,2}^{2,0}\left(\frac{C(g+1)\gamma_{th}}{\overline{\gamma}_{RF}^2}\bigg|\begin{matrix}-\\1,0\end{matrix}\right)\left[1 - \frac{\xi^2 2^{\alpha+\beta-3}}{\pi\Gamma(\alpha)\Gamma(\beta)}G_{4,9}^{7,2}\left(\frac{(\alpha\beta\kappa)^2 C(k+1)\gamma_{th}}{16\overline{\gamma}_{FSO}\overline{\gamma}_{RF}}\bigg|\begin{matrix}\psi_1\\\psi_2\end{matrix}\right)\right]$$

$$P_e = \frac{1}{2}\left\{1 - \sum_{k=0}^{N-1}\sum_{t=0}^{M-1}\sum_{u=0}^{t}\sum_{k_1=0}^{t}\sum_{k_2=0}^{k_1}\sum_{n=0}^{\infty}\Omega\binom{t}{k_1}\binom{k_1}{k_2}X_0^{t-k_1}\left(Y_n^{(k_1-k_2)}*Z_n^{(k_2)}\right)\frac{(1/\overline{\gamma}_{FSO})^{\beth}}{\left(1+\frac{k+u+1}{\overline{\gamma}_{RF}}\right)^{1+\beth}}\left[\Gamma(1+\beth) - \right.\right. \tag{35}$$

$$\frac{\xi^2 2^{\alpha+\beta-3}}{\pi\Gamma(\alpha)\Gamma(\beta)}G_{5,9}^{7,3}\left(\frac{(\alpha\beta\kappa)^2 C(k+1)}{16\overline{\gamma}_{FSO}(\overline{\gamma}_{RF}+k+u+1)}\bigg|\begin{matrix}-\beth,\psi_1\\\psi_2\end{matrix}\right)\right] - \sum_{k=0}^{N-1}\sum_{t=0}^{M-1}\sum_{u=0}^{t}\sum_{k_1=0}^{t}\sum_{k_2=0}^{k_1}\sum_{n=0}^{\infty}\Omega\binom{t}{k_1}\binom{k_1}{k_2}X_0^{t-k_1}\left(Y_n^{(k_1-k_2)}*\right.$$

$$\left.Z_n^{(k_2)}\right)\frac{(1/\overline{\gamma}_{FSO})^{\beth}}{\left(1+\frac{k+u+1}{\overline{\gamma}_{RF}}\right)^{1+\beth}}G_{1,2}^{2,1}\left(\frac{C(k+1)}{\overline{\gamma}_{RF}(\overline{\gamma}_{RF}+k+u+1)}\bigg|\begin{matrix}-\beth\\1,0\end{matrix}\right) + \sum_{k=0}^{N-1}\sum_{g=0}^{N-1}\sum_{t=0}^{M-1}\sum_{u=0}^{t}\sum_{k_1=0}^{t}\sum_{k_2=0}^{k_1}\sum_{n=0}^{\infty}\Omega\times$$

$$\frac{(-1)^g N}{g+1}\binom{N-1}{g}\binom{t}{k_1}\binom{k_1}{k_2}X_0^{t-k_1}\left(Y_n^{(k_1-k_2)}*Z_n^{(k_2)}\right)\frac{(1/\overline{\gamma}_{FSO})^{\beth}}{\left(1+\frac{k+g+u+2}{\overline{\gamma}_{RF}}\right)^{1+\beth}}\left[G_{1,2}^{2,1}\left(\frac{C(k+1)}{\overline{\gamma}_{RF}(\overline{\gamma}_{RF}+k+g+u+2)}\bigg|\begin{matrix}-\beth\\1,0\end{matrix}\right) - \right.$$

$$\left.\left.\frac{\xi^2 2^{\alpha+\beta-3}}{\pi\Gamma(\alpha)\Gamma(\beta)}G_{1,0:0,2:4,9}^{1,0:2,0:7,2}\left(\begin{matrix}1+\beth\\-\end{matrix}\bigg|\begin{matrix}-\\1,0\end{matrix}\bigg|\begin{matrix}\psi_1\\\psi_2\end{matrix}\bigg|\frac{C(k+1)}{\overline{\gamma}_{RF}(\overline{\gamma}_{RF}+k+g+u+2)}, \frac{(\alpha\beta\kappa)^2 C(g+1)}{16\overline{\gamma}_{FSO}(\overline{\gamma}_{RF}+k+g+u+2)}\right)\right]\right\}$$

integrate the multiplication of only two Meijer-G functions and one exponential function; this integration has an available answer.

Using (41) from Appendix A, the linear equivalent of $\left(\frac{\xi^2}{\Gamma(\alpha)\Gamma(\beta)} G_{2,4}^{3,1}\left(\alpha\beta\kappa\sqrt{\frac{\gamma}{\bar{\gamma}_{FSO}}}\bigg|\begin{matrix}1,\xi^2+1\\\xi^2,\alpha,\beta,0\end{matrix}\right)\right)^t$ becomes as $\left(X_0\left(\frac{\gamma_{th}}{\bar{\gamma}_{FSO}}\right)^{\frac{\xi^2}{2}} + \sum_{n=0}^{\infty} Y_n \left(\frac{\gamma_{th}}{\bar{\gamma}_{FSO}}\right)^{\frac{n+\alpha}{2}} + \sum_{n=0}^{\infty} Z_n \left(\frac{\gamma_{th}}{\bar{\gamma}_{FSO}}\right)^{\frac{n+\beta}{2}}\right)^t$. By using trinomial expansion it becomes as follows:

$$\sum_{k_1=0}^{t}\sum_{k_2=0}^{k_1}\binom{t}{k_1}\binom{k_1}{k_2}\left(X_0\left(\frac{\gamma_{th}}{\bar{\gamma}_{FSO}}\right)^{\frac{\xi^2}{2}}\right)^{t-k_1} \times \left(\sum_{n=0}^{\infty}Y_n\left(\frac{\gamma_{th}}{\bar{\gamma}_{FSO}}\right)^{\frac{n+\alpha}{2}}\right)^{k_1-k_2}\left(\sum_{n=0}^{\infty}Z_n\left(\frac{\gamma_{th}}{\bar{\gamma}_{FSO}}\right)^{\frac{n+\beta}{2}}\right)^{k_2}. \quad (32)$$

By using [53, Eq.0.314] and [53, Eq.0.316], (32) becomes as:

$$\sum_{k_1=0}^{t}\sum_{k_2=0}^{k_1}\sum_{n=0}^{\infty}\binom{t}{k_1}\binom{k_1}{k_2}X_0^{t-k_1}\left(Y_n^{(k_1-k_2)}*Z_n^{(k_2)}\right)\left(\frac{\gamma_{th}}{\bar{\gamma}_{FSO}}\right)^{\beth}, \quad (33)$$

where $\beth = \frac{n+\xi^2(t-k_1)+\alpha(k_1-k_2)+\beta k_2}{2}$, and (*) denotes the convolution and the subscript $h_n^{(k)}$ means that $h_n$ is convolved $(k-1)$ times with itself.

By substituting equivalent of $\left(\frac{\xi^2}{\Gamma(\alpha)\Gamma(\beta)} \times G_{2,4}^{3,1}\left(\alpha\beta\kappa\sqrt{\frac{\gamma}{\bar{\gamma}_{FSO}}}\bigg|\begin{matrix}1,\xi^2+1\\\xi^2,\alpha,\beta,0\end{matrix}\right)\right)^t$ from (33) into (27), Outage Probability of the proposed structure in Gamma-Gamma atmospheric turbulence with the effect of pointing errors becomes equal to (34).

By substituting (34) into (30) and using [52, Eq. 07.34.21.0081.01] and [52, Eq. 07.34.21.0088.01], BER of DPSK modulation in Gamma-Gamma atmospheric turbulence with the effect of pointing error becomes equal to (35), where $G_{-:-:-}^{-:-:-}\left(\bar{-}\bigg|\bar{-}\bigg|\bar{-}\bigg|.,.\right)$ is the Extended Bivariate Meijer-G function [54].

It has worth to ask why should investigate complex multi-hop structures. Why should derive such complicated expressions? Aren't they expandable from of single and dual hop structures? The answer is "No", it is necessary to investigate multi-hop structures independently. Because effects of number of relays can't be shown while expanding single and dual hop structures. Actually number of relays affect the decision made on the signal and the system performance.

### 5.2 Negative Exponential atmospheric turbulence

Substituting (29) into (30), and substituting Meijer-G equivalent of $e^{-\lambda v\sqrt{\frac{\gamma}{\bar{\gamma}_{FSO}}}}$, BER of DPSK modulation in Negative Exponential atmospheric turbulence becomes as (36). This integral cannot be solved for the same reason as (31), because of multiplication of three Meijer-G functions and one exponential function. Furthermore, the trick used to obtain exact solution for (31) does not work here. Therefore, this paper tried to solve (36) asymptotically. The idea of this paper for solving this integral is to obtain a linear substitution for CDF of Negative Exponential in (36). In fact, after this substitution, for solving (36), it is sufficient to integrate the multiplication of only two Meijer-G functions and one exponential function; this integration has an available answer.

By substituting the CDF of Negative Exponential atmospheric turbulence from Appendix B, (12), (22) and (24) into (25), and using binomial expansion of $\left[1-\theta\gamma^{\frac{1}{2}}\left(1-e^{-\frac{\gamma_{th}}{\bar{\gamma}_{RF}}}\right)\right]^{M-1}$ as $\sum_{t=0}^{M-1}\sum_{u=0}^{t}\binom{M-1}{t}\binom{t}{u}(-1)^{t+u}e^{-\frac{u\gamma_{th}}{\bar{\gamma}_{RF}}}\left(\theta\gamma_{th}^{\frac{1}{2}}\right)^t$, asymptotic Outage Probability of the proposed structure in Negative Exponential atmospheric turbulence becomes as (37).

$$P_e = \frac{1}{2}\int_0^{\infty} e^{-\gamma}\Bigg\{1 - \sum_{k=0}^{N-1}\sum_{t=0}^{M-1}\sum_{u=0}^{t}\frac{\Lambda}{\sqrt{\pi}}e^{-\frac{(k+u+1)\gamma}{\bar{\gamma}_{RF}}}G_{0,2}^{2,0}\left(\frac{(\lambda v)^2\gamma}{4\bar{\gamma}_{FSO}}\bigg|\begin{matrix}-\\0,1/2\end{matrix}\right)G_{0,2}^{2,0}\left(\frac{\gamma C(k+1)}{\bar{\gamma}_{RF}^2}\bigg|\begin{matrix}-\\1,0\end{matrix}\right) - \sum_{k=0}^{N-1}\sum_{t=0}^{M-1}\sum_{u=0}^{t}\frac{\Lambda}{\sqrt{\pi}}e^{-\frac{(k+u+1)\gamma}{\bar{\gamma}_{RF}}}G_{0,3}^{3,0}\left(\frac{\lambda^2\gamma C(k+1)}{4\bar{\gamma}_{FSO}\bar{\gamma}_{RF}}\bigg|\begin{matrix}-\\1,0,1/2\end{matrix}\right) + \sum_{k=0}^{N-1}\sum_{g=0}^{N-1}\sum_{t=0}^{M-1}\sum_{u=0}^{t}\binom{N-1}{g}(-1)^g \times \frac{N\Lambda}{\pi(g+1)}e^{-\frac{(k+g+u+2)\gamma}{\bar{\gamma}_{RF}}}G_{0,2}^{2,0}\left(\frac{(\lambda v)^2\gamma}{4\bar{\gamma}_{FSO}}\bigg|\begin{matrix}-\\0,1/2\end{matrix}\right)G_{0,2}^{2,0}\left(\frac{\gamma C(k+1)}{\bar{\gamma}_{RF}^2}\bigg|\begin{matrix}-\\1,0\end{matrix}\right)G_{0,3}^{3,0}\left(\frac{\lambda^2\gamma C(g+1)}{4\bar{\gamma}_{FSO}\bar{\gamma}_{RF}}\bigg|\begin{matrix}-\\1,0,1/2\end{matrix}\right)\Bigg\}d\gamma \quad (36)$$

$$P_{out}(\gamma_{th}) \cong 1 - \Bigg(\sum_{k=0}^{N-1}\binom{N-1}{k}\binom{M-1}{t}\binom{t}{u}\frac{(-1)^{(k+t+u)}N}{\sqrt{\pi}(k+1)}\left(\theta\gamma_{th}^{\frac{1}{2}}\right)^t e^{-\frac{(k+u+1)\gamma_{th}}{\bar{\gamma}_{RF}}}G_{0,3}^{3,0}\left(\frac{\lambda^2 C(k+1)\gamma_{th}}{4\bar{\gamma}_{FSO}\bar{\gamma}_{RF}}\bigg|\begin{matrix}-\\1,0,\frac{1}{2}\end{matrix}\right) + \sum_{k=0}^{N-1}\binom{N-1}{k}\binom{M-1}{t}\binom{t}{u}\frac{(-1)^{(k+t+u)}N}{k+1}\left(\theta\gamma_{th}^{\frac{1}{2}}\right)^t e^{-\frac{(k+u+1)\gamma_{th}}{\bar{\gamma}_{RF}}}G_{0,2}^{2,0}\left(\frac{C(k+1)\gamma_{th}}{\bar{\gamma}_{RF}^2}\bigg|\begin{matrix}-\\1,0\end{matrix}\right) - \sum_{k=0}^{N-1}\sum_{g=0}^{N-1}\binom{N-1}{k} \times \binom{N-1}{g}\binom{M-1}{t}\binom{t}{u}\frac{(-1)^{(k+g+t+u)}N}{\sqrt{\pi}(g+1)(k+1)}\left(\theta\gamma_{th}^{\frac{1}{2}}\right)^t e^{-\frac{(k+g+u+2)\gamma_{th}}{\bar{\gamma}_{RF}}}G_{0,2}^{2,0}\left(\frac{C(k+1)\gamma_{th}}{\bar{\gamma}_{RF}^2}\bigg|\begin{matrix}-\\1,0\end{matrix}\right)G_{0,3}^{3,0}\left(\frac{\lambda^2 C(g+1)\gamma_{th}}{4\bar{\gamma}_{FSO}\bar{\gamma}_{RF}}\bigg|\begin{matrix}-\\1,0,\frac{1}{2}\end{matrix}\right)\Bigg). \quad (37)$$

$$P_e \cong \frac{1}{2}\Bigg\{1 - \sum_{k=0}^{N-1}\sum_{t=0}^{M-1}\sum_{u=0}^{t}\varsigma\,\theta^t\frac{1}{\left(1+\frac{k+u+1}{\bar{\gamma}_{RF}}\right)^{\frac{t}{2}+1}}G_{1,2}^{2,1}\left(\frac{C(k+1)}{\bar{\gamma}_{RF}(\bar{\gamma}_{RF}+k+u+1)}\bigg|\begin{matrix}-\frac{t}{2}\\1,0\end{matrix}\right) - \sum_{k=0}^{N-1}\sum_{t=0}^{M-1}\sum_{u=0}^{t}\frac{\varsigma\,\theta^t}{\sqrt{\pi}} \times \frac{1}{\left(1+\frac{k+u+1}{\bar{\gamma}_{RF}}\right)^{\frac{t}{2}+1}}G_{1,3}^{3,1}\left(\frac{\lambda^2 C(k+1)}{4\bar{\gamma}_{FSO}(\bar{\gamma}_{RF}+k+u+1)}\bigg|\begin{matrix}-\frac{t}{2}\\1,0,1/2\end{matrix}\right) + \sum_{k=0}^{N-1}\sum_{g=0}^{N-1}\sum_{t=0}^{M-1}\sum_{u=0}^{t}\binom{N-1}{g}(-1)^g\frac{N\varsigma\,\theta^t}{\sqrt{\pi}(g+1)} \times \frac{1}{\left(1+\frac{k+g+u+2}{\bar{\gamma}_{RF}}\right)^{\frac{t}{2}+1}}G_{1,0:0,2:0,3}^{1,0:2,0:3,0}\left(1+\frac{t}{2}\bigg|\begin{matrix}-\\1,0\end{matrix}\bigg|\begin{matrix}-\\1,0,1/2\end{matrix}\bigg|\frac{C(k+1)}{\bar{\gamma}_{RF}(\bar{\gamma}_{RF}+k+g+u+2)},\frac{\lambda^2 C(g+1)}{4\bar{\gamma}_{FSO}(\bar{\gamma}_{RF}+k+g+u+2)}\right)\Bigg\} \quad (38)$$

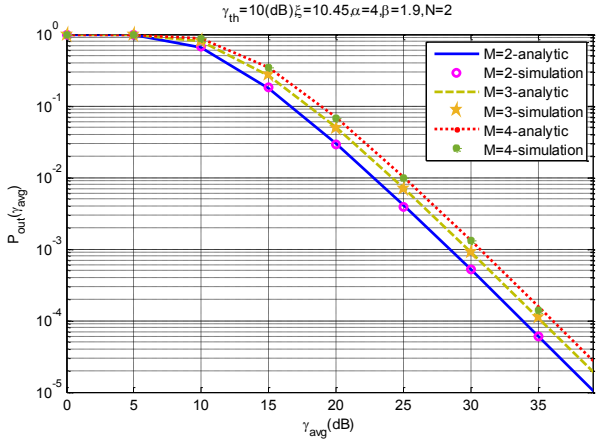

**Fig. 3:** Outage Probability of the proposed structure as a function of average SNR for various number of relays for moderate regime of Gamma-Gamma atmospheric turbulence with the effect of pointing error when number of receive antennas is $N = 2$ and $\gamma_{th} = 10dB$.

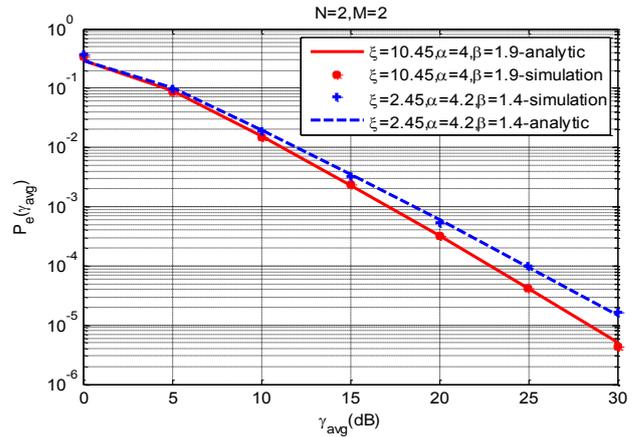

**Fig. 5:** Bit Error Rate of the proposed structure as a function of average SNR, for moderate and strong regimes of Gamma-Gamma atmospheric turbulence with the effect of pointing error, when number of relays is $M = 2$ and number of antennas is $N = 2$.

By substituting (37) into (30), and using [52, Eq. 07.34.21.0081.01] and [52, Eq. 07.34.21.0088.01], asymptotic BER of DPSK modulation in Negative Exponential atmospheric turbulence becomes equal to (38), where $\varsigma = \binom{N-1}{k}\binom{M-1}{t}\binom{t}{u} \times (-1)^{k+t+u}\frac{N}{k+1}$.

Though with modern computing tools, it is now easier to produce analytical expressions, it should be considered that even modern computing tools cannot produce closed-form expression shorter than these formulations; because Meijer-G function is the shortest possible expression that could be found for mathematical expressions. Meijer-G is a frequently used mathematical tool in FSO system for deriving performance investigation expressions. This function has a complex structure; it is not easy to have insight about it; furthermore, the proposed multi-hop structure is complicated; therefore, the complexity of the derived mathematical formulations is reasonable. Many publications in FSO system performance used Meijer-G function. Though they did not have mathematical insights, they provided enough physical insights from figure plots in their results section.

## 6. Comparison between analytic and simulation results

In this section, the obtained analytical results are compared with MATLAB simulations. It is assumed that FSO and RF links, have equal average SNRs ($\gamma_{avg} = \bar{\gamma}_{FSO} = \bar{\gamma}_{RF}$). For simplicity and without loss of generality it is assumed that $\eta = 1$ and $C = 1$ [55]. The proposed structure is evaluated at different atmospheric turbulence intensities as well as different number of receive antennas ($N$) and relays ($M$). The threshold outage SNR of the proposed structure and is denoted by $\gamma_{th}$. The parameters $\alpha, \beta,$ and $\xi$ in Gamma-Gamma atmospheric turbulence with the effect of pointing error are assumed to be $\alpha = 4, \beta = 1.9, \xi = 10.45$ in moderate regime and $\alpha = 4.2, \beta = 1.4, \xi = 2.45$ in strong regime.

### 6.1 Gamma-Gamma atmospheric turbulence with the effect of pointing error

The effect of number of relays and number of receive antennas on the performance of the proposed structure are respectively investigated in Fig. 3, and Fig. 4. It is indicated in Fig. 3 that the Outage Probability increases while increasing number of relays.

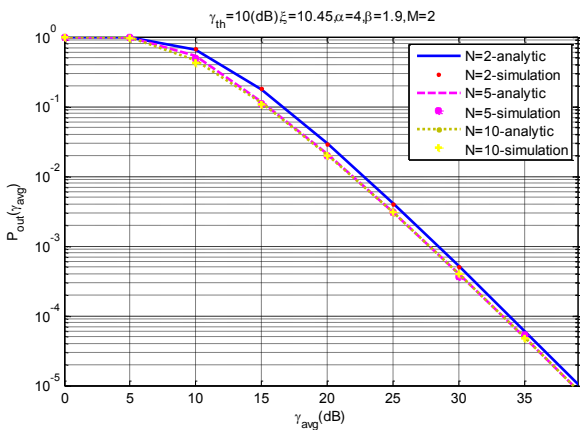

**Fig. 4:** Outage Probability of the proposed structure, as a function of average SNR for various number of receive antennas for moderate regime of Gamma-Gamma atmospheric turbulence with the effect of pointing error, when number of relays is $M = 2$ and $\gamma_{th} = 10dB$.

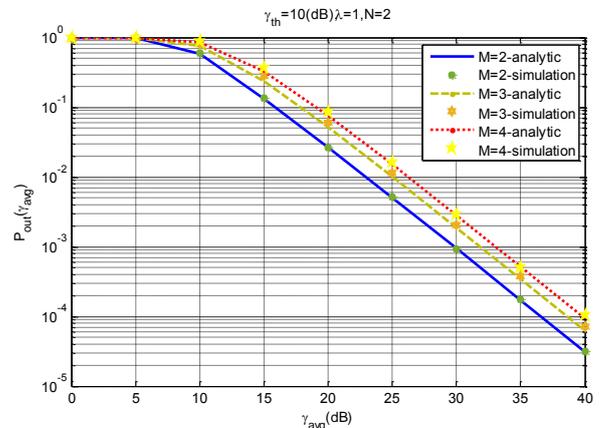

**Fig. 6:** Outage Probability of the proposed structure as a function of average SNR for various number of relays for Negative Exponential atmospheric turbulence with unit variance, and $\gamma_{th} = 10dB$.

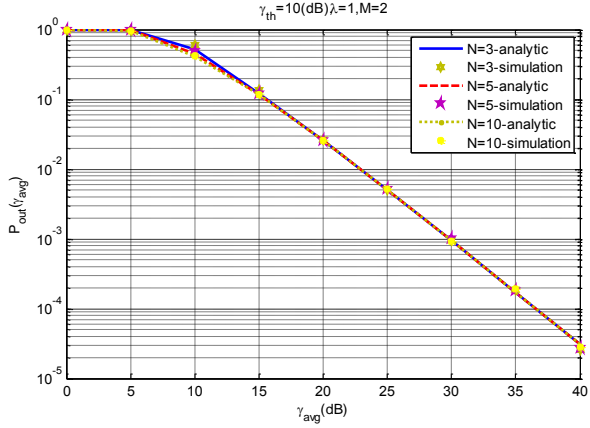 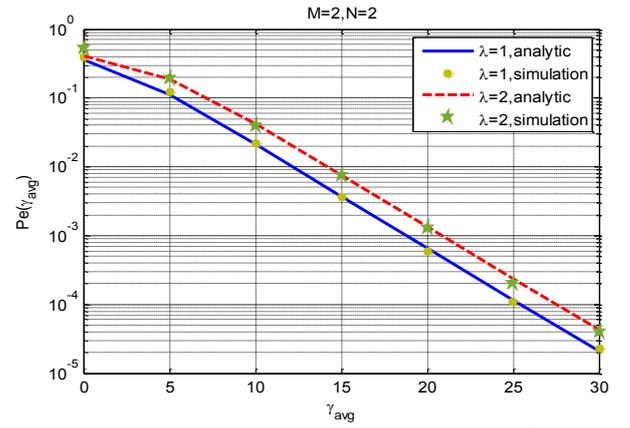

**Fig. 7:** Outage Probability of the proposed structure as a function of average SNR for various number of receive antennas, for Negative Exponential atmospheric turbulence with unit variance, when number of relay is $M = 2$ and $\gamma_{th} = 10dB$.

**Fig. 8:** Bit Error Rate of the proposed structure as a function of average SNR, for various variances of Negative Exponential atmospheric turbulence, when number of relays is $M = 2$ and number of receiving antenna is $N = 2$.

Because in the proposed structure, the outage occurs when the instantaneous SNR of each of the individual relays becomes lower than a threshold. In series relay structure, increasing number of relays is equivalent to increasing number of decisions made on the signal, which is equivalent to increasing probability of making wrong decisions that leads to system performance degradation. But in parallel relay structure, because of diversity, increasing the number of relays decreases the Outage Probability.

Also, it can be seen in Fig. 3 that at wide range of target $P_{out}$, $\gamma_{avg}$ difference between cases with different number of relays is fixed, meaning that by increasing number of relays, a constant fraction of consuming power should be added in order to maintain performance of the system. Hence, the proposed structure does not require additional processing to adjust this fraction adaptively. So this system is cost effective and suitable for mobile communications.

Fig. 4 shows that the Outage Probability is almost independent of number of receive antennas. The proposed structure has the same performance at different number of receive antennas. Therefore, in the proposed structure, number of receive antennas is not important, which means that the proposed structure provides a favorable performance with low power consumption and low complexity. However, at low $\gamma_{avg}$, there is little difference between cases with different number of antennas, which is because of the noise dominant effect at this range.

Fig. 5, compares BER of DPSK modulation in moderate and strong regimes of Gamma-Gamma atmospheric turbulence with the effect of pointing error. As can be seen, performance of the proposed structure has little dependence on the intensity of Gamma-Gamma atmospheric turbulence with the effect of pointing error. For example $\gamma_{avg}$ difference between moderate and strong regimes, at $P_e = 10^{-4}$ is about 2dB and at $P_e = 10^{-3}$ is about $1.5dB$. This is an advantage for the proposed structure, because identical performance at different atmospheric turbulence intensities means that it is not required to consume much more power or use additional processing to maintain performance when atmospheric turbulence changes.

*6.2 Negative Exponential atmospheric turbulence*

Fig. 6, and Fig. 7 respectively investigate the effects of number of relays and number of antennas on system Outage Probability. As can be seen in Fig. 6, at various target $P_{out}$, $\gamma_{avg}$ difference between cases with different number of relays is fixed. Actually series relay structures show a trade-off between performance and capacity; in the sense that increasing number of relays degrades system performance, but increases the system capacity. Fig. 7 shows little $\gamma_{avg}$ difference between cases with different number of receive antennas. Therefore, the same as moderate regimes, system performance in saturate regime is independent of number of receive antennas.

In Fig. 8, BER performance of DPSK in various variances of Negative Exponential atmospheric turbulence are compared. As can be seen, obtained asymptotic results match with the simulations at $\gamma_{avg} \geq 5\text{dB}$. Compared with the moderate regime, system performance at saturate regime is a bit more dependent on atmospheric turbulence intensity. However, the same as moderate regime, system performs favourable even at low $\gamma_{avg}$ in saturate regime.

## 7. Conclusion

In this paper, a novel multi-hop hybrid FSO / RF structure is presented as a solution for long range mobile communications. This structure is made of two main parts; at the first part a fixed gain amplify and forward relay connects mobile user and source base station in a long range link. In order to have better performance, relay uses multiple receive antennas with selection combining scheme. At the second part a multi-hop hybrid parallel FSO / RF system with demodulate and forward relaying connects source and destination base stations. In order to have better performance, opportunistic selection is performed at each relay. Considering FSO link at Gamma-Gamma atmospheric turbulence with the effect of pointing error in moderate to strong regimes, and at Negative Exponential atmospheric turbulence in saturate regime, new closed form exact and asymptotic expressions are derived for BER and Outage Probability of the proposed structure. Derived expressions are verified by MATLAB simulations. Results indicate that performance of proposed structure is almost independent of number of receive antennas. Also performance difference between cases with different number of relays is fixed. The proposed structure shows favourable performance at wide range of atmospheric turbulences form moderate to saturate regimes. According to these results, the proposed structure is affordable in terms of cost, power, and complexity; also it is particularly suitable for long range mobile communications.

This paper aimed to present a new comprehensive structure as a solution for long range communications as one of the most challenging problems in communication systems. Results of this paper indicated that at wide range of atmospheric turbulences even with considering the effect of pointing errors, it is possible to have a reliable connection even with low power consumption. This paper showed that there is no need to "do" implement complicated coding or detection techniques or use heavy processing or massive antennas to make communication reliable. Complexity is a non-dissociable part of most of the existing communication structures. Because they should serve many users with high reliability, data rate and favorable performance, which cannot be provided without complexity. The main point of this paper is that only by adding some simple relays, it's possible to have favorable performance at wide range of atmospheric turbulences even with the effect of pointing error, even at low power consumption at a long range link. In this structure the mobile user should not consume more power or add complexity, also the receiver should not implement additional processing or complicated detection techniques; so this is the main physical insight that could be provided from proposed complex mathematical exercises.

## Appendix A

### Exact CDF of Gamma-Gamma atmospheric turbulence with the effect of pointing error

Using [52, Eq.07.34.26.0004.01], the pdf of Gamma-Gamma atmospheric turbulence with the effect of pointing error becomes equal to:

$$f_{\gamma_{FSO}}(\gamma) = \frac{\xi^2 \Gamma(\alpha-\xi^2)\Gamma(\beta-\xi^2)}{2\Gamma(\alpha)\Gamma(\beta)\gamma}\left(\alpha\beta\kappa\sqrt{\frac{\gamma}{\bar\gamma_{FSO}}}\right)^{\xi^2} {}_1F_2\left(0; 1-\alpha+\xi^2, 1-\beta+\xi^2; \alpha\beta\kappa\sqrt{\frac{\gamma}{\bar\gamma_{FSO}}}\right) + \frac{\xi^2 \Gamma(\xi^2-\alpha)\Gamma(\beta-\alpha)}{2\Gamma(\alpha)\Gamma(\beta)\Gamma(\xi^2+1-\alpha)\gamma}\left(\alpha\beta\kappa\sqrt{\frac{\gamma}{\bar\gamma_{FSO}}}\right)^{\alpha} {}_1F_2\left(\alpha-\xi^2; 1-\xi^2+\alpha, 1-\beta+\alpha; \alpha\beta\kappa\sqrt{\frac{\gamma}{\bar\gamma_{FSO}}}\right) + \frac{\xi^2 \Gamma(\alpha-\beta)\Gamma(\xi^2-\beta)}{2\Gamma(\alpha)\Gamma(\beta)\Gamma(\xi^2+1-\beta)\gamma}\left(\alpha\beta\kappa\sqrt{\frac{\gamma}{\bar\gamma_{FSO}}}\right)^{\beta} {}_1F_2\left(\beta-\xi^2; 1-\xi^2+\beta, 1-\alpha+\beta; \alpha\beta\kappa\sqrt{\frac{\gamma}{\bar\gamma_{FSO}}}\right),$$
(39)

where ${}_pF_q(a_1,...,a_p; b_1,...,b_q; z)$ is the Hyper-geometric function [52, Eq. 07.31.02.0001.01]. Using [52, Eq. 07.23.02.0001.01], the above expression becomes equal to:

$$f_{\gamma_{FSO}}(\gamma) = \frac{\xi^2}{2\bar\gamma_{FSO}} X_0 \left(\frac{\gamma}{\bar\gamma_{FSO}}\right)^{\frac{\xi^2}{2}-1} + \sum_{n=0}^{\infty} \frac{n+\alpha}{2\bar\gamma_{FSO}} Y_n \left(\frac{\gamma}{\bar\gamma_{FSO}}\right)^{\frac{n+\alpha}{2}-1} + \sum_{n=0}^{\infty} \frac{n+\beta}{2\bar\gamma_{FSO}} Z_n \left(\frac{\gamma}{\bar\gamma_{FSO}}\right)^{\frac{n+\beta}{2}-1},$$
(40)

where $Y_n = \frac{\xi^2 \Gamma(\xi^2-\alpha)\Gamma(\beta-\alpha)(\alpha-\xi^2)_n}{(n+\alpha)\Gamma(\alpha)\Gamma(\beta)\Gamma(\xi^2+1-\alpha)(1-\xi^2+\alpha)_n(1-\beta+\alpha)_n n!}(\alpha\beta\kappa)^{n+\alpha}$,
$Z_n = \frac{\xi^2 \Gamma(\alpha-\beta)\Gamma(\xi^2-\beta)(\beta-\xi^2)_n}{(n+\beta)\Gamma(\alpha)\Gamma(\beta)\Gamma(\xi^2+1-\beta)(1-\xi^2+\beta)_n(1-\alpha+\beta)_n n!}(\alpha\beta\kappa)^{n+\beta}$, and
$X_0 = \frac{\Gamma(\alpha-\xi^2)\Gamma(\beta-\xi^2)}{\Gamma(\alpha)\Gamma(\beta)}(\alpha\beta\kappa)^{\xi^2}$. Note that $(.)_n$ is the pochhammer symbol [51]. Integrating (34), the CDF of Gamma-Gamma atmospheric turbulence with the effect of pointing error becomes equal to[1]:

$$F_{\gamma_{FSO}}(\gamma) = X_0 \left(\frac{\gamma}{\bar\gamma_{FSO}}\right)^{\frac{\xi^2}{2}} + \sum_{n=0}^{\infty} Y_n \left(\frac{\gamma}{\bar\gamma_{FSO}}\right)^{\frac{n+\alpha}{2}} + \sum_{n=0}^{\infty} Z_n \left(\frac{\gamma}{\bar\gamma_{FSO}}\right)^{\frac{n+\beta}{2}}.$$
(41)

---

[1]. The main idea of such transformation from Meijer-G to linear summation is taken from [51], Eq. (10)-(14). In this paper, no truncated value of n has revealed to show the convergence of the infinite series, because use of a truncated value of n does not lead to an exact result, and "=" sign should be changed to "≃". But in MATLAB implementation, n taken in (0, 20) works.

## Appendix B

### Asymptotic CDF of Negative Exponential Atmospheric Turbulence

By using [52, Eq. 07.34.06.0006.01] the asymptotic CDF of Negative Exponential atmospheric turbulence becomes as follows:

$$F_\gamma(\gamma) = 1 - e^{-\lambda\sqrt{\frac{\gamma}{\bar\gamma_{FSO}}}} = 1 - \frac{1}{\sqrt{\pi}} G_{0,2}^{2,0}\left(\frac{\lambda^2 \gamma}{\bar\gamma_{FSO}} \Big| \frac{-}{0,\frac{1}{2}}\right) \cong -\Gamma\left(-\frac{1}{2}\right)\left(\frac{\lambda^2}{4\pi\bar\gamma_{FSO}}\right)^{\frac{1}{2}} \gamma^{\frac{1}{2}} = \theta \gamma^{\frac{1}{2}},$$
(42)

where $\theta = -\Gamma\left(-\frac{1}{2}\right)\left(\frac{\lambda^2}{4\pi\bar\gamma_{FSO}}\right)^{\frac{1}{2}}$.